\documentclass[12pt] {article}
\usepackage{amsmath} 
\usepackage{amssymb} 
\usepackage{epsfig}
\newcommand{\vph}{\boldsymbol{\varphi}}
\newcommand{\vrh}{\boldsymbol{\rho}}
\newcommand{\val}{{\boldsymbol{\alpha}}}

\newcommand{\vaa}{\boldsymbol{a}}
\newcommand{\veta}{\boldsymbol{\eta}}
\newcommand{\vom}{\boldsymbol{\omega}}

\newcommand{\vP}{{\bf P}}
\newcommand{\vQ}{{\bf Q}}
\newcommand{\bfe}{{\bf e}}
\newcommand{\mbar}{{\overline m}}

\makeatletter

\@addtoreset{equation}{section}
\makeatother

\newcommand{\limi}[1]{\raisebox{-0.23cm}{~\shortstack{ $\mbox{lim
}$ \\
${\vspace{-0.2cm} _{#1}}$}}}

\def\cc#1{\kern .7em\hfill #1 \hfill\kern .7em}

\newcommand{\nc}{\newcommand}
\nc{\beqa}{\begin{eqnarray}}
\nc{\eeqa}{\end{eqnarray}}
\def\agt{
\mathrel{\raise.3ex\hbox{$>$}\mkern-14mu\lower0.6ex\hbox{$\sim$}}
}
\def\alt{
\mathrel{\raise.3ex\hbox{$<$}\mkern-14mu\lower0.6ex\hbox{$\sim$}}
}

\begin {document}
\bibliographystyle{unsrt}    

\begin{flushright}APCTP-2000-020\\KIAS-P01005\end{flushright}
\vspace{1cm}
\centerline{\Large \bf Perturbative test of exact vacuum expectation values}
\centerline{\Large \bf of local fields in affine Toda theories}
\vskip 1cm
\centerline{\large Changrim Ahn$^{1,2}$, P.\ Baseilhac$^3$, 
Chanju Kim$^{4}$ and Chaiho Rim$^{5,2}$}
\vskip 1cm
\centerline{\it$^{1}$Department of Physics, Ewha Womans University}
\centerline{\it Seoul 120-750, Korea}
\vskip .4cm
\centerline{\it$^{2}$Asia Pacific Center for Theoretical Physics}
\centerline{\it Yoksam-dong 678-39, Seoul, 135-080, Korea}
\vskip .4cm
\centerline{\it$^{3}$Department of Mathematics, University of York}
\centerline{\it Heslington, York YO105DD, United Kingdom}
\vskip .4cm
\centerline{\it $^{4}$ School of Physics, Korea Institute for Advanced Study}
\centerline{\it Seoul, 130-012, Korea}
\vskip .4cm
\centerline{\it $^{5}$ Department of Physics, Chonbuk National University}
\centerline{\it Chonju 561-756, Korea}
\vskip 0.7cm
\centerline{\small PACS: 11.25.Hf, 11.55.Ds}
\vskip 0.7cm

\begin{abstract}
Vacuum expectation values of local fields for all dual pairs of non-simply
laced affine Toda field theories recently proposed are checked against
perturbative analysis. The computations based on Feynman diagram
expansion are performed upto two-loops. Agreement is obtained.
\end{abstract}

\newpage

\pagestyle{plain} 

\setcounter{page}{2}
\section{Introduction}
The vacuum expectation values (VEV)s of local fields play an important
 role in quantum field theory (QFT) and statistical mechanics
\cite{1,2}. In QFT defined as perturbed conformal field theory (CFT),
they constitute the basic ingredients for multipoint
correlation functions, using short-distance expansions \cite{2,3}. 
Some times ago, important progress was made in 
the calculations of the VEVs in two dimensional integrable QFT. In
 ref. \cite{4}, an explicit expression for the VEVs of the exponential
 field in the sine-Gordon and sinh-Gordon models - $A_1^{(1)}$ affine
Toda field theory (ATFT) - was proposed. Moreover, it was
 shown in \cite{Fateev20} that this expression can be obtained as the
 minimal solution of certain ``reflection relations'' which involve the
 Liouville ``reflection amplitude'' \cite{Zam0}, where the
 sinh-Gordon QFT was considered as a perturbed Liouville conformal field
theory. Later, this ``reflection relations'' method was successfully generalized
to other models, for which the VEVs were calculated. We refer the reader
to refs. \cite{Fateev2,8,9,10,Bas} for details.

Among the family of known integrable QFTs, it was thus natural to study 
the case of dual pairs of non-simply laced
ATFTs, beyond the simply laced one which had been previously considered
 \cite{Fateev4}. Beside the technical
aspect, such VEVs can provide interesting informations as
this class of models appears in various physics contexts
\cite{tr,zN,vays,muss}. For instance, in \cite{Fateev4} the following
applications were studied : particular correlation functions in a
special case of 3-D $U(1)$ or XY model, VEV of the spin field $\sigma$
in the ${\mathbb{Z}}_n$-Ising models \cite{zamfat} perturbed by the leading thermal
operator, asymptotics of the cylindrically symmetric solutions of the
classical Toda equations. More recently, exact off-shell results for coupled
minimal models were considered in \cite{coupled}.

ATFTs can be considered as perturbed Toda field theories (TFT)s. In
\cite{non} the ``reflection amplitudes'' for all non-simply laced Toda field
 theories (TFT)s were proposed as well as the exact relation between the masses
of the particles and the parameters in the action associated with ATFTs. On the one hand,
 reflection amplitudes are the main objects which can be
used for studying the UV asymptotics of the ground state energy $E(R)$
(or effective central charge $c_{eff}(R)$) for the system on the circle
of size $R$ \cite{Zam0,hid,kim,non}. In particular, the comparison of its asymptotics
at small $R$ with the same quantity which can be independently
calculated from the $S$-matrix data using TBA method \cite{YY,Za} can be
considered as a non-trivial test for the $S$-matrix amplitudes \cite{non} proposed
in \cite{DGZ,CDS}.
On the other hand, reflection amplitudes were also used 
to calculate the exact VEVs for all dual pairs of non-simply laced ATFTs
 \cite{non}.

However, in order to support the exact VEVs, various checks are
desirable. The main reason is that the ``reflection relations'' method has
{\it no} mathematical proof yet. For the two simplest cases (sinh-Gordon and
Bullough-Dodd), the exact VEVs have been checked non-perturbatively but
also perturbatively using the
standard perturbation theory \cite{4,Fateev2}, the perturbation
theory in the radial \cite{pog1}
or angular quantization \cite{pog2} framework. The purpose of this paper is to check
 the VEVs conjectures of general affine Toda theories
using standard perturbative calculations upto two-loops. 
In the next section, we recall some basic facts about ATFTs and 
the axiomatic equations satisfied by the VEVs which lead to the
exact solutions given in \cite{non}. 
Perturbative analysis follows in sect.3 where we carefully compute $<\vph>$ 
perturbatively upto one-loop and also the VEVs of some composite operators 
upto two-loops.

\section{Exact vacuum expectation values in affine Toda field theories}
Let us first recall some known results about ATFTs which are relevant in further analysis.
The ATFT with real coupling $b$ corresponding to the affine Lie algebra\,\footnote{Throughout
the paper, we denote an untwisted algebra as $\widehat{\cal G}$, while $\widehat{\cal
G}^\vee$ refers to a twisted one. Furthermore, $\cal G$ denotes a
finite Lie algebra.} $\widehat{\cal G}$ is generally described by the action in the Euclidean space :
\beqa
{\cal A} =  \int d^2x \Big[\frac{1}{8\pi}(\partial_\mu\vph)^2 + 
\sum_{i=0}^{r}\mu_{\bfe_i}e^{b\bfe_i\cdot \vph}\Big],\label{action}
\eeqa
where $\{\bfe_i\}\in\Phi_{\bf s}$ $(i=1,...,r)$ 
is the set of simple roots of $\widehat{\cal G}$ of rank $r$ and $-\bfe_0$ is a maximal root satisfying 
\beqa
\bfe_0+\sum_{i=1}^{r}n_i\bfe_i=0.
\eeqa
The fields in (\ref{action}) are normalized such that :
\beqa
<\varphi_a(x)\varphi_b(y)>=-\delta_{ab}\log|x-y|^2.
\eeqa
For the simply laced case, since all vertex operators in the potential
 possess the same conformal dimension they all renormalize in the
 same way. It is then sufficient to introduce one scale
 parameter\,\footnote{For the sinh-Gordon model ($A_1^{(1)}$ ATFT) $\mu$ is generally 
called the cosmological constant.} $\mu$ in action (\ref{action}).  
However, for the non-simply laced case (except $BC_r\equiv A_{2r}^{(2)}$ - 
$r\geq 2$ - affine Lie algebra in
which case three different parameters are necessary) 
we have to introduce two different
parameters\,\footnote{We choose the convention that the length squared of the long 
roots are four for $C_r^{(1)}$ and two for the other untwisted
algebras.} : one is associated 
with the set of standard roots of length $|\bfe_i|^2=2$ and 
is denoted by $\mu_{\bfe_i}=\mu$ whereas
the other,  denoted by $\mu_{\bfe_i}=\mu'$,  is associated with the set
of non-standard roots of length $|\bfe_i|^2=l^2 \neq 2$.

The ATFTs can be considered as perturbed CFTs. Without the term
with the zeroth root $\bfe_0$, the action in (\ref{action})
describes a TFT which is conformal. To do it, one introduce a charge to
 infinity defined by : 
\beqa
{\bf Q} =   b\vrh   +   \frac{1}{b}\vrh^\vee \ \ \ \ \ \ \ \mbox{where}
\ \ \
\ \ \ \ \ \vrh=\frac{1}{2}\sum_{\val>0} \val\ \ \ \ \mbox{and}\ \ \ \ 
\vrh^\vee=\frac{1}{2}\sum_{\val>0} \val^\vee                                \eeqa
are respectively the Weyl and dual Weyl vector of $\cal G$. The sums in
their definitions run over all positive roots $\{\val\}\in\Phi_+$,
 dual roots $\{\val^\vee\}\in\Phi_+^\vee$. Then, the stress-energy tensor $T(z)$, where
 $z=x_1+ix_2$, $\overline{z}=x_1-ix_2$ are complex coordinates of ${\mathbb R}^2$,  
\beqa
T(z)=-\frac{1}{2}(\partial_z\vph)^2 + {\bf Q}\cdot \partial_z^2\vph
\eeqa
ensures the local conformal invariance of the TFT. 
The corresponding central charges were calculated in \cite{Hol}.
Defining ${\vaa}=(a_1,...,a_r)$, the exponential fields 
\beqa
V_{\vaa}(x) = \exp({\vaa} \cdot {\bf \vph})(x)\label{vop}
\eeqa
are spinless conformal primary fields with dimensions :
\beqa
\Delta({\vaa})= \frac{{\bf Q}^2}{2}-\frac{({\vaa}-{\bf Q})^2}{2}.
\eeqa
By analogy with the Liouville field theory \cite{Curt,Neveu,Zam0} the physical space of states
${\cal H}$ in  the TFTs consists of the continuum variety of primary states
 associated with the exponential fields (\ref{vop}) and their conformal descendents with :
\beqa
{\vaa}=i{\bf P} + {\bf Q}\ \ \ \mbox{and}\ \ \ \vP\in{\mathbb R}^r.\label{defa}
\eeqa

Besides the conformal invariance TFTs possess an extended
symmetry generated by $W({\cal G})$-algebra \cite{FL}. Indeed, for any arbitrary
Weyl group element $\hat s\in{\cal W}$  the fields
$V_{{\bf Q}+{\hat s}({\vaa-{\bf Q}})}(x)$ are  reflection images of each other
and are related by the linear transformation : 
\beqa
V_{{\vaa}}(x) = R_{{\hat s}}({\vaa})V_{{\bf Q}+{\hat s}({\vaa-{\bf Q}})}(x)\label{refl}
\eeqa
where $R_{{\hat s}}({\vaa})$ is called the ``reflection amplitude'', an
important object in CFT which 
defines the two-point functions of the operator $V_{\vaa}$.
 In \cite{non} the following expression for the reflection amplitude 
$R_{\hat s}({\vaa})$ for non-simply laced TFT was obtained : 
\beqa
R_{\hat s}({\vaa}) = \frac{A_{{\hat s}i{\bf P}}}{A_{i{\bf P}}}\label{R}
\eeqa
where 
\beqa
A_{i{\bf P}} && \equiv A({\bf P})\nonumber \\
&&=\ 
\prod_{i=1}^{r}[\pi\mu_{\bfe_i}\gamma(\bfe_i^2b^2/2)]^{i\vom^\vee_i\cdot\vP/b}
\times \prod_{\val>0} \Gamma(1-i{\bf P} \cdot \val b) 
\Gamma(1-i{\bf P} \cdot \val^\vee/b)
\nonumber
\eeqa
with (\ref{defa}), the fundamental co-weights $\vom_i^\vee$ and we denote 
$\gamma(x)=\Gamma(x)/\Gamma(1-x)$ as usual. We accept (\ref{R}) as the 
proper analytical continuation of the function $R_{\hat s}({\vaa})$ for all
${\vaa}$. For $\hat{s}_i\in{\cal W}_{\bf s}$, the subset of Weyl group 
elements associated with the simple roots $\bfe_i$, notice that the ratio
$A(\hat{s}_i\vP)/A(\vP)$ reduce to the reflection amplitude 
$S_L(\bfe_i, \vP)$ of the Liouville field theory \cite{Zam0} :
\beqa \label{Rj}
\frac{A(\hat{s}_i\vP)}{A(\vP)} &=& S_L(\bfe_i, \vP) \nonumber \\
  &=& [\pi\mu_{\bfe_i}\gamma(\bfe_i^2b^2/2)]^{-i\vP \cdot \bfe_i^\vee/b}
       \frac{\Gamma(1+i\vP \cdot \bfe_ib)
	\Gamma(1+i\vP\cdot\bfe_i^\vee/b)}
        {\Gamma(1-i\vP\cdot\bfe_ib)\Gamma(1-i\vP\cdot\bfe_i^\vee/b)}.
\eeqa

Then, as ATFTs can be realized as CFTs perturbed by some
relevant operators \cite{Zam1}, in the conformal perturbation theory
(CPT) approach one can formally rewrite any $N$-point correlation
functions of local operators ${\cal O}_a(x)$ as :
\beqa
<{\cal O}_{a_1}(x_1)...{\cal O}_{a_N}(x_N)>_{ATFT}=Z^{-1}(\lambda)
<{\cal O}_{a_1}(x_1)...{\cal O}_{a_N}(x_N)e^{-\lambda\int d^2x\Phi_{pert}(x)}>_{TFT}\nonumber
\eeqa
where 
\beqa
Z(\lambda)=<e^{-\lambda\int d^2x\Phi_{pert}(x)}>_{TFT},\nonumber
\eeqa
$\Phi_{pert}$ is the perturbing local field, $\lambda$ is
the CPT expansion parameter which characterizes the strength of the
perturbation and \ $<...>_{TFT}$ denotes the expectation value in the
TFT.  Whereas vertex operators (\ref{vop}) satisfy reflection relations
(\ref{refl}) in the CFT, the CPT
framework provides\,\footnote{At the moment, there is no {\it rigorous}
proof of this assumption.} similar relations among their expectation values
in the perturbed case.
In other words, if dots stands for any local insertion one has :
\beqa
<V_{\vaa}(x)(...)>_{TFT}= R_{{\hat s}}
({\vaa})<V_{{\bf Q}+{\hat s}({\vaa-{\bf Q}})}(x)(...)>_{TFT}.\label{arb}
\eeqa
Then, if we define the one-point function $G(\vaa)$
as the VEV of the vertex operator $V_{\vaa}(x)$ for non-simply laced
ATFT by : 
\begin{eqnarray}
G({\vaa}) = <\exp({\vaa}\cdot\vph)(x)>_{ATFT}
\label{VEV1}
\end{eqnarray}
one can formally rewrite this expression\,\footnote{In fact, the
integrals in (\ref{expd}) are highly infrared divergent. By analogy with
the situation appearing in the perturbed Liouville QFT \cite{Fateev2}, one
can get around this infrared problem by considering a 2D world-sheet
$\Sigma_g$ topologically equivalent to a sphere equipped by a background
metric $g_{\nu\sigma}(x)=\rho(x)\delta_{\nu\sigma}$. Then the
terms $\rho(y_k)$ which appear in the integrals analogous to those in
(\ref{expd}) provide an efficient infrared cut-off. We report the reader
to \cite{Fateev2} for details.} as : 
\beqa
<\!e^{{\vaa}\cdot\vph}(x)\!>_{ATFT}\!=\!Z^{-1}(\lambda)\sum_{n=0}^{\infty}\!\frac{(-\lambda)^n}{n!}
\!\!\int\! \prod_{j=1}^{n} d^2 y_j
<\!e^{{\vaa}\cdot\vph}(x)e^{b{\bfe_0}\cdot\vph}(y_1)...
e^{b{\bfe_0}\cdot\vph}(y_n)\!>_{TFT}.\label{expd}
\eeqa
Indeed, using (\ref{arb}) one expects that similar relations hold
for $G(\vaa)$. If this VEV satisfies the system of functional equations
associated with ${\cal W}_{\bf s}$ then it also
automatically satisfies more complicated reflection relations.
Furthermore, as was shown in previous works \cite{zN,vays}, ATFTs can
be understood as different perturbation of TFTs. The simplest
case (beyond the sinh-Gordon model) is the Bullough-Dodd model which can be
understood alternatively \cite{Fateev2} as
a perturbed Liouville CFT with coupling constant $b$ {\it or}\ \ a perturbed
 Liouville CFT with coupling constant $-b/2$. Here one can
proceed similarly. We denote $\Phi_{\bf s}(\cal G)$ as the set of simple
roots of the finite
Lie algebra $\cal G$, $\veta$ the extra-root associated with the
perturbation and $\{\epsilon_i\}$ an orthogonal basis \ \
($\epsilon_i\cdot\epsilon_j=\delta_{ij}$)\ \ in ${\mathbb R}^r$. Each one of
 the ATFT Lagrangian representation, denoted ${\cal L}_{b}\big[\Phi_{\bf s}({\cal G})\big]$, associated with $\Phi_{\bf s}({\cal G})$ and the coupling constant $b$ can be
rewrite in two different ways :
\beqa
\ \ \ \ \ \ \ \ \ \ \ \ \ \ \ \ \ \ \bullet\ \ \ {\cal L}_b\big[\Phi_{\bf s}(B_r^{(1)})\big]&\equiv& {\cal L}_{b}\big[\Phi_{\bf s}(B_r)\ \oplus \ \veta\equiv
 \bfe_0 = -(\epsilon_1+\epsilon_2)\big],\nonumber\\
&\equiv& {\cal L}_{-b}\big[{\overline \Phi_{\bf s}(D_r)}\ \oplus \ \veta\equiv
-\epsilon_r\big];\nonumber\\
&& \nonumber\\
\bullet\ \ \ {\cal L}_{b}\big[\Phi_{\bf s}(C_r^{(1)})\big]&\equiv& {\cal L}_{b}\big[\Phi_{\bf s}(C_r)\ \oplus \ \veta\equiv
 \bfe_0 = -2\epsilon_1\big],\nonumber\\
&\equiv& {\cal L}_{-b}\big[{\overline \Phi_{\bf s}(C_r)}\ \oplus \ \veta\equiv
-2\epsilon_r\big];\nonumber\\
&& \nonumber\\
\bullet\ \ \ {\cal L}_b\big[\Phi_{\bf s}(F_4^{(1)})\big]&\equiv& {\cal L}_b\big[\Phi_{\bf s}(F_4)\ \oplus
\ \veta\equiv \bfe_0 = - \epsilon_1-\epsilon_2\big],\nonumber\\
&\equiv& {\cal L}_{-b}\big[{\overline \Phi}_{\bf s}(B_4)\ \oplus \ \veta\equiv
 -\frac{1}{2}(\epsilon_1-\epsilon_2-\epsilon_3-\epsilon_4)\big];\nonumber\\
&& \nonumber\\
\bullet\ \ \ {\cal L}_b\big[\Phi_{\bf s}(G_2^{(1)})\big]&\equiv&{\cal L}_b\big[\Phi_{\bf s}(G_2)\ \oplus \ \veta\equiv
 \bfe_0 = -\sqrt{2}\epsilon_1\big],\nonumber\\
&\equiv& {\cal L}_{-b}\big[{\overline \Phi_{\bf s}(A_2)}\ \oplus\ \veta\equiv
-{\sqrt {2/3}}\epsilon_2\big].\nonumber
\eeqa
where the different sets of simple roots are reported in Appendix B.
Using (\ref{arb}), it implies that the VEV (\ref{VEV1})
must satisfy {\it simultaneously} two irreducible systems of functional equations
corresponding to two different sets ${\cal W}_{\bf s}$. 
It results that $G(\vaa)$ obeys the
 functional equations 
\beqa
G({\tau\vaa}) =  S_L(\bfe_j,\vP)G(\tau({\bf Q}+{\hat s}_j({\vaa-{\bf Q}}))) \ \ \ \ \mbox{for all}\ \ \
\ {\hat s}_j\in{\cal W}_{\bf s}\label{funtau}
\eeqa
where \\
\\
$\bullet$ $B_r^{(1)}$ : $(\tau)_{ij}=\delta_{ij}$\ \ \ for ${\cal G}\equiv B_r$\
\ \ and \ \ \  $(\tau)_{ij}=-\delta_{i\ r+1-j}$\ \ \ for ${\cal G}\equiv D_r$;\\
$\bullet$ $C_r^{(1)}$ : $(\tau)_{ij}=\delta_{ij}$
\ \ \ \ \ \ \  \ \ \ \ \ \ \ \ \ \ \ \ and \ \ \  $(\tau)_{ij}=-\delta_{i\ r+1-j}$\ \ \ for
${\cal G}\equiv C_r$;\\
$\bullet$ $F_4^{(1)}$ : $(\tau)_{ij}=\delta_{ij}$\ \ \ for ${\cal G}\equiv F_4$\
\ \ and \ \ \ $(\tau)_{ij}=\delta_{ij}(\delta_{2j}+\delta_{3j}+\delta_{4j}-\delta_{1j})$
\ \ \ for ${\cal G}\equiv B_4$;\\
$\bullet$ $G_2^{(1)}$ : $(\tau)_{ij}=\delta_{ij}$\ \ \ for ${\cal G}\equiv
G_2$\ \
\  and \ \ \  $(\tau)_{ij}=-\delta_{i\ 3-j}$\ \ for ${\cal G}\equiv A_2$\\
\\
with coupling constant $b$. Notice that by simply looking at the
Dynkin diagram symmetry of $B_r^{(1)}$ and $C_r^{(1)}$ - see figure 1 - one can also
differently deduce :
\beqa
&&G(a_1,a_2,...,a_{r-1},a_r)=G(-a_1,a_2,...,a_{r-1},a_r)\  \ \ \ \ \ \ \ \ \ \ \mbox{for} \ \ \
B_r^{(1)}\ ; \label{symet}\\
&&G(a_1,a_2,...,a_{r-1},a_r)=G(-a_r,-a_{r-1},...,-a_2,-a_1)\  \ \ \ \mbox{for} \ \ \ C_r^{(1)}.\nonumber
\eeqa
The reflection relations (\ref{funtau}) (or, equivalently the relations
(\ref{symet}) for $B_r^{(1)}$ and $C_r^{(1)}$)
 constituted the starting point in deriving the expectation values
$G(\vaa)$. Following previous works, we also assumed that $G(\vaa)$ is a meromorphic
function of $\vaa$.

\vspace{8mm}
\centerline{\epsfig{file=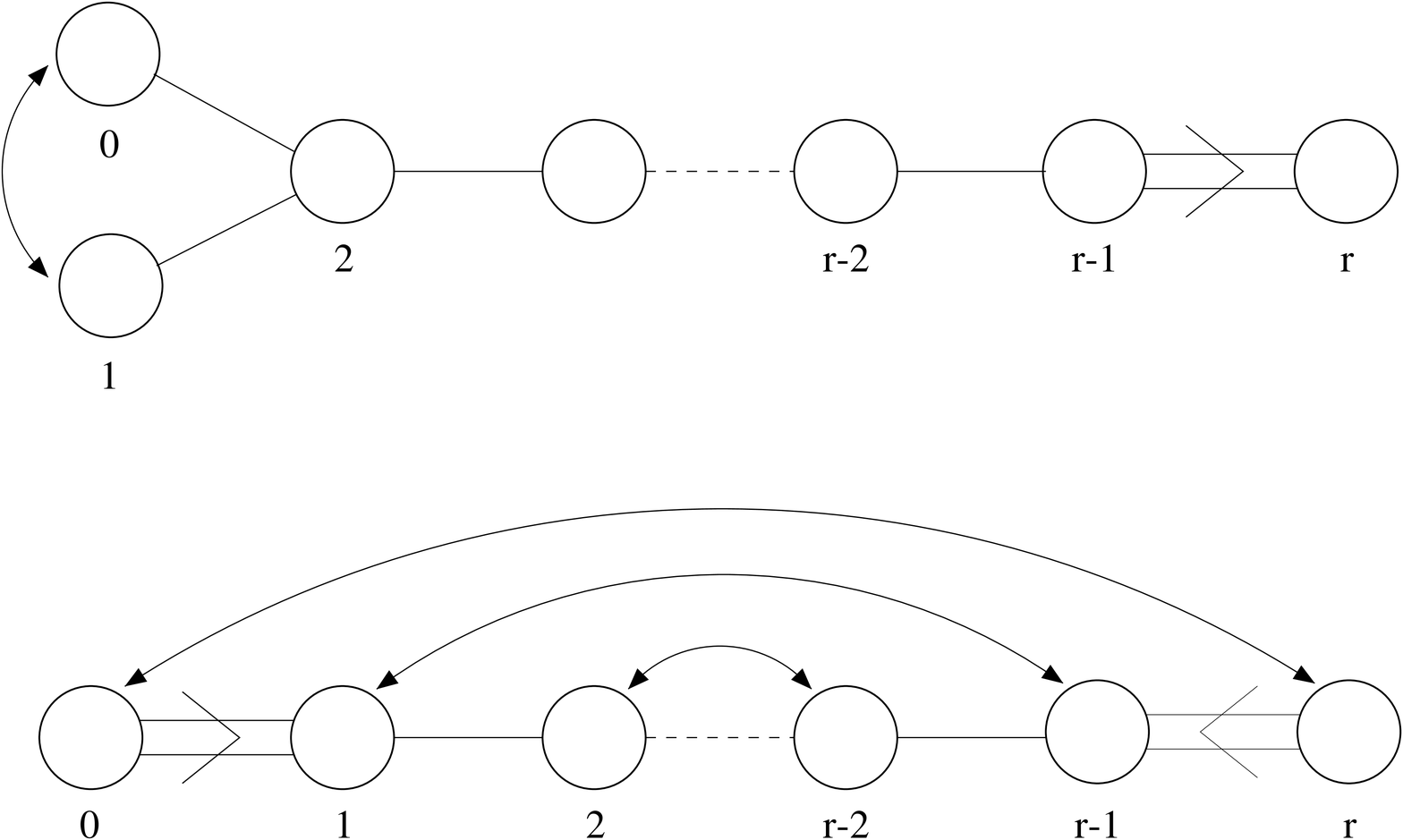,height=45mm,width=80mm}}
\vspace{1mm}
\begin{center}
{\small \underline{Figure 1}  -  Automorphisms associated with the Dynkin diagram}\\
{\small \ \ $B_r^{(1)}$ and $C_r^{(1)}$ corresponding to (\ref{symet}).}
\end{center}

Furthermore, for real coupling constant $b$, the spectrum for any dual pair of non-simply laced
ATFT consists of $r$ particles with the masses $M_a$ ($a=1,...,r$)
expressed in terms of the mass parameter $\mbar$. These spectra are reported in appendix
 A.  The exact relation between the parameters of the action $\mu$ and $\mu'$
and the masses associated with the spectrum of the physical particles was
 obtained in \cite{non} using the Bethe ansatz method (see for example
 \cite{mumass,fateev}). We report the reader to \cite{non} for details.
 By replacing these mass-$\mu$ relations in the ``minimal'' 
solution\,\footnote{Notice that the prefactor which was given
in ref. \cite{non} was presented in a slightly different, but
equivalent, form.} of the functional equations (\ref{funtau}), 
 the following exact expression for the VEVs (\ref{VEV1}) was proposed
\cite{non} :
\beqa
G({\vaa}) &=& 
\Big[\mbar k({\cal G})\kappa({\cal G})\Big]^{-{\vaa}^2}\Big[\frac{\mu\gamma(1+b^2)}
{\mu'\gamma
(1+b^2l^2/2)}\Big]^{\frac{{\bf d}.{\vaa}(1-B)}{Hb}}\Big[\frac{\big(-\pi\mu\gamma(1+b^2)
\big)^{l^2/2}}{-\pi\mu'\gamma(1+b^2l^2/2)}\Big]^{\frac{{\bf d}.{\vaa}B}{Hb}}\nonumber \\
&&\ \ \ \ \ \ \ \ \ \ \ \ \ \ \ \ \ \ \ \times \exp \int_{0}^{\infty} \frac{dt}{t} 
\Big( {\vaa}^2 e^{-2t} - {\cal F}({\vaa},t) \Big)\label{Gafin}           
\eeqa
with
\beqa
{\cal F}({\vaa},t) = \sum_{{\val} >0} \Big[\ \frac{\sinh({a_{\val}}bt)
 \sinh\big(({a_{\val}}b -2{Q_{\val}}b + H(1+b^2))t\big)
\sinh(\big(\frac{b^2|\val|^2}
{2}+1\big)t)}{\sinh (t)\ \sinh(\frac{b^2|\val|^2}{2}t)\ \sinh(H(1+b^2)t)}\ 
\Big]\nonumber 
\eeqa
where we denote $a_{\val}=\vaa\cdot\val$ and 
\beqa
{\bf d}=\frac{\vrh^\vee h^\vee - \vrh h}{1-l^2/2}.\nonumber
\eeqa
The expressions $k({\cal G})$ and $\kappa({\cal G})$ can be found in
\cite{non}. Here, it is convenient to introduce the ``deformed'' Coxeter number \cite{DGZ,CDS} :
\beqa
H= h(1-B) + h^\vee B \ \ \ \ \ \ \ \ \mbox{with}\ \ \ \ \ \ B=\frac{b^2}{1+b^2}
\eeqa
where $h$\ (resp. $h^\vee$) is the Coxeter (resp. dual Coxeter) number of $\cal
G$ (resp. ${\cal G}^\vee$). The integral in (\ref{Gafin}) is convergent iff :
\beqa
{\val\cdot \bf Q}-H(b+1/b)\ < \ {\mathfrak R}e(\val\cdot\vaa)\ 
<\ {\val\cdot \bf Q}\ \ \ \ \ \ \ \mbox{for all}\ \ \ \ \ \ \ \val\in\Phi_{+}
\eeqa
and is defined through analytic continuation outside this domain.
Particular case of (\ref{Gafin}) corresponds to the simply laced one
for which the result is in perfect agreement with \cite{Fateev4}.

Similarly, it is straightforward to obtain the VEVs of an ATFT based on
a twisted affine Lie algebra ${\widehat{\cal G}}^\vee$. The reflection
amplitudes corresponding to the TFT, i.e. the conformal part 
were easily obtained from (\ref{R}) by using the duality relation
 for the parameters $\mu_{\bfe_i}$ and $\mu_{\bfe_i}^\vee$ 
associated with the dual pairs of ATFTs \cite{non} : 
\beqa
\pi\mu_{\bfe_i}\gamma\big(\frac{b^2\bfe_i^2}{2}\big)\
 =\ \Big[\pi\mu_{\bfe_i}^\vee\gamma\big(\frac{{\bfe_i^\vee}^2}{2b^2}\big)
\big]^{b^2\bfe_i^2/2} \label{duale} 
\eeqa
and the change \ $b \rightarrow 1/b$. Each one of the Lagrangian associated
with ${\widehat{\cal G}}^\vee$ can be written in two different ways. In any case, the
resulting system of functional equations which has to be satisfied by
the VEV is nothing else than the dual of (\ref{funtau}). To express the
corresponding solution in terms of the mass of the physical particles,
 the mass-$\mu$ relations in the twisted case \cite{non} are used. Finally,
 the result for the VEV $G({\vaa})$ for all twisted 
affine Lie algebras is obtained from (\ref{Gafin}) with the change
 \ $b \rightarrow 1/b$.

It is similarly straightforward to study the $BC_r\equiv
A_{2r}^{(2)}$ (self-dual) remaining case which was considered in \cite{moi}.  

Notice that the expectation values (\ref{Gafin}) can be used to derive the
 bulk free energy of the ATFT :
\beqa
f_{\widehat{\cal G}} =  -\limi{V\rightarrow
\infty}\frac{1}{V}\ln Z, 
\eeqa
where $V$ is the volume of the 2D space and $Z$ is the singular part 
of the partition function associated with action (\ref{action}).
For specific values ${\vaa} \in b\{\bfe_i\}$, with
$\{\bfe_i\}\in\Phi_{\bf s}$ ($i=1,...,r$) or $\bfe_0$, the integral
 in (\ref{Gafin}) can be evaluated explicitely.
Using the exact mass-$\mu$ relations and the obvious relations:
\beqa
\partial_{\mu} f(\mu)=\sum_{\{i\}}< e^{b\bfe_i\cdot\vph}>\ \ \ \ \ \ \mbox{or}
\ \ \ \ \ \ \partial_{\mu'} f(\mu')=\sum_{\{i'\}}< e^{b\bfe_{i'}\cdot\vph}>
\eeqa
where $\{i\}$ and $\{i'\}$ denotes respectively the whole set of long and
short roots, the following bulk free energy was obtained \cite{non} :
\beqa
f_{\widehat{\cal G}}&=&\frac{\mbar^2 \sin(\pi/H)}{8\sin(\pi B/H)\sin(\pi(1-B)/H)}\,,
  \ \ \  \ \ \ \  \ \ \ \ \ \ 
     \qquad {\widehat{\cal G}}=B_r^{(1)}\ \mbox{and} \ C_r^{(1)},\nonumber\\
f_{\widehat{\cal G}}&=&\frac{\mbar^2 \cos(\pi(1/3-1/H))}{16\cos(\pi/6)\sin(\pi B/H)
                                           \sin(\pi(1-B)/H)},
     \qquad {\widehat{\cal G}}=G_2^{(1)}\ \mbox{and} \ F_4^{(1)}\nonumber
\eeqa
and similarly with the change $B\rightarrow (1-B)$ for
$(B_r^{(1)})^{\vee}$, $(C_r^{(1)})^{\vee}$, $(G_2^{(1)})^{\vee}$, and
$(F_4^{(1)})^{\vee}$. In particular, these results were in perfect
agreement with the values obtained using the Bethe ansatz approach \cite{non}.

\section{Perturbative checks}

To support the result (\ref{Gafin}) of \cite{non} beyond the non-perturbative check
(provided by the bulk free energy calculation), we present here a
 perturbative check. We expand the vacuum expectation value (\ref{Gafin}) 
in power series in $b$ and compare  each coefficient
with the one obtained from standard Feynman perturbation theory
associated with (\ref{action}). In the  first part of this section,
we consider the VEV of the field $<\vph>$ which is given by :
\beqa
 <{\bf \vph} > =\frac { \delta }{ \delta \vaa} G(\vaa )|_{\vaa = 0}\, .
\label{vevphi} 
\eeqa
Since the result renders the same conclusion for all ATFTs,
we choose $D_r^{(1)}$ series as illustrative examples and omit the details 
for other simply laced cases ($A_r^{(1)}$ case is trivial as seen shortly). 
It also provides a useful step to the calculation of $B_r^{(1)}$ series 
which is obtained from $D_r^{(1)}$ through folding procedure. 
Finally we present the result of an exceptional algebra $G_2^{(1)}$.
 
In a second part, as an additional check we also consider the
``fully connected'' composite operator expectation value of
$<\vph^a\vph^b>$ defined by 
\beqa
<<\vph^a \vph^b>> &\equiv& <\vph^a \vph^b> -<\vph^a><\vph^b>
= \frac{1}{2} \frac{\delta^2 \ln G(\vaa)}{\delta \vaa^a \delta \vaa^b}\label{vevcomp}
\left.\right|_{\vaa =0}\label{comp}
\eeqa
Since these quantities are quite complicated to calculate perturbatively,
we will content ourselves with considering only some simple combinations 
of them upto two loops for $B_3^{(1)}$, $C_2^{(1)}$ and $G_2^{(1)}$ cases.  

\vspace{0.3cm}
\subsection{Perturbative checks of \ $<\vph>$}

Using (\ref{Gafin}) and (\ref{vevphi}) one finds the result :
%
%
%
\beqa
<\vph>\! &=& \!\frac{{\bf d}}{Hb}\ln\Big[\frac{\mu\gamma(1+b^2)}{\mu'\gamma(1+b^2l^2/2)}\Big]
+ B \frac{{\bf d}}{Hb}(l^2/2-1)\ln\big[-\pi\mu\gamma(1+b^2)\big]
\nonumber\\
&+&\! b \int_0^{\infty}\! dt \sum_{{\bf \val}>0} \Big[\val
\frac{\sinh\big((2{\val\cdot\vQ}b - H(1+b^2))t\big)
\sinh(\big(\frac{b^2|\val|^2}{2}+1\big)t)}
{\sinh(t)\sinh (H(1+b^2)t)\sinh(\frac{b^2|\val|^2}{2}t)}\Big]\,.\label{moy} 
\eeqa
To proceed further\,\footnote{However, notice that $<\vph>=0$ 
identically for $A_r^{(1)}$ series since $\mu' =\mu$, $l^2 =2$ and \ $\sum_{{\bf \val}>0} \val\ \ 
\sinh\left( (2{\val \cdot \vQ} - hQ) bt \right)  =0$
 \cite{destri,Fateev4}.}, we expand 
$<\vph>$ order by order in $b$ and write the result as : 
\beqa
< \vph> = {1 \over b} {\cal K} + b {\cal L} 
+ {\cal O}(b^2)\ .\label{se} 
\eeqa

For the simply laced case, this expression is drastically simplified :
\beqa
&&{\cal K} = 
  - \sum_{\val >0}  \val \ \  
\ln \gamma \left( 
{ \val \cdot \vrh \over h }
\right) \,; 
\\
&&
{\cal L} =
- \int_0^\infty dt \, {\coth (t)  \over \sinh (h t)} 
\left\{ \sum_{\val >0} 
\val\ \sinh (( h - 2 \val \cdot \vrh) t) 
\right\}\ .\nonumber 
\eeqa
Let us introduce the component notation :
${\cal K}_i = \ \bfe _i \cdot {\cal K}$ and 
${\cal L}_i =  \bfe _i \cdot {\cal L}$.

For the $D_r^{(1)}$ series, the non-perturbative results are given as
\beqa
&& {\cal K}_1= {\cal K}_{r-1} = {\cal K}_r 
= - ( 1 - { 4 \over h} ) \ln 2
\label{simply-power1}\ ;\\
&&{\cal L}_1 = {\cal L}_{r-1} = {\cal L}_r
={1 \over 2h} \left( 
\xi ( {1 \over h}) +
\xi ( {2 \over h} ) - \xi ( {1 \over 2 } + {1 \over h}) 
-\xi ( {1 \over 2})
\right)\nonumber
\end{eqnarray}
where $h=2r-2$ and  for $  k = 2, 3, \cdots, r-2 $ we have 
\begin{eqnarray}
&&{\cal K}_k = \ln 2 - (1 - {4 \over h} ) \ln 2\,;
\label{simply-power2}\\
&&{\cal L}_k = 
{1 \over 2h} \left(
 -\ \xi ({ 1 \over 2} + {k \over h}) 
 - \xi ({1 \over 2} + { k-1 \over h})
 -\xi ( {k \over h} )
 - \xi ({k-1 \over h})
 \right.
 \nonumber \\
 && \qquad \qquad  
 \left.
 + \ \ \xi ({2k -2 \over h} )
 +2 \xi ({2k -1 \over h})
 + \xi ({2k \over h} )
\right)
\nonumber
\eeqa
where we define $\xi (x) = \Psi(x) + \Psi (1 -x) $ in terms of 
the di-gamma function,   $\Psi (x) =  d \ln \Gamma(x) / dx$.

Perturbative analysis of the action (\ref{action}) 
begins with shifting  
$\vph \to \vph_{cl} + \vph $ such that 
$\vph_{cl}$ satisfies the minimum of the ATFT potential.
This classical solution reproduces exactly the leading term in 
(\ref{se}) :
\beqa
\vph_{cl} \cdot \bfe_i = {\cal K}_i \,. 
\eeqa
This identity provides the amusing relations
among the $\gamma(x)$-functions, when $x$ is related to Lie algebra 
quantity, which is observed for general case in  ref. \cite{Fateev4,non}

One-loop perturbative calculation is conveniently done 
using the classical mass eigenstate representation \cite{CDS}.
$D_r^{(1)}$ series representation is given by : 
\begin{eqnarray}
&&\bfe_1  = ( - l_1^1, -l_1^2, \cdots, -l_1^{r-2},1, 0)
\label{D-mass}\\
&&\bfe_k  = ( l_{k-1}^1- l_k^1, l_{k-1}^2-l_k^2, \cdots, 
l_{k-1}^{r-2} -l_k^{r-2}, 0, 0)
\ \ \mbox{ for } \ k =2, \cdots, r-2 
\nonumber\\
&&\bfe_{r-1}  = (  l_{r-2}^1, l_{r-2}^2, 
\cdots, l_{r-2}^{r-2}, 0, -1)
\nonumber\\
&&\bfe_r  = ( l_{r-2}^1, l_{r-2}^2, 
\cdots, l_{r-2}^{r-2}, 0, 1)
\nonumber
\end{eqnarray}
where $l_k^a = \frac{2}{\sqrt{h}} \sin\frac{2 ak\pi}{h}$.

The next-to leading order term, i.e. the field expectation value to 
the one-loop order, $< \vph >_b$
is given by tad-pole diagrams 
which in general needs to be appropriatly regularized. 
The perturbative result is, however, finite for 
the mass eigenstate representation and 
does not depend on the regularization scheme for $D_r^{(1)}$ series (and
in general for simply laced cases).

To distinguish from the component notation,  $\varphi_j$, which
 is obtained from the proposed VEV, 
the perturbative  mass eigenstate component is 
denoted as $\Phi^c$. 
If the values are correct, then 
the relation between these two quantities should be
$ \varphi_j =\sum_c  \Phi^c  \bfe_j^c$ 
where $\bfe_j^c$ is the $c$-th  
component of the mass eigenstate representation $\bfe_j$.

$\Phi^c$ vanishes when 
$c =r-1,\ r$ and  $c=$ odd $\le r-2$, 
\begin{equation}
<\Phi^c >_b = 0\,;
\end{equation}
and otherwise
\begin{eqnarray} \label{diagram}
<\Phi^c >_b &=& \frac12\left[
\begin{picture}(50,50)(-25,0)
\put(0,0){\circle{40}}
\put(0,-20){\line(0,-1){10}}
\put(-5,-25){\line(1,-1){10}}
\put(-5,-35){\line(1,1){10}}
\put(0,-43){$c$}
\put(0,27){$\frac{c}{2}$}
\end{picture}
+
\begin{picture}(50,50)(-25,0)
\put(0,0){\circle{40}}
\put(0,-20){\line(0,-1){10}}
\put(-5,-25){\line(1,-1){10}}
\put(-5,-35){\line(1,1){10}}
\put(0,-43){$c$}
\put(0,27){$\frac{h-c}{2}$}
\end{picture}
+
\begin{picture}(50,50)(-25,0)
\put(0,0){\circle{40}}
\put(0,-20){\line(0,-1){10}}
\put(-5,-25){\line(1,-1){10}}
\put(-5,-35){\line(1,1){10}}
\put(0,-43){$c$}
\put(0,25){$r$}
\end{picture}
+
\begin{picture}(50,50)(-25,0)
\put(0,0){\circle{40}}
\put(0,-20){\line(0,-1){10}}
\put(-5,-25){\line(1,-1){10}}
\put(-5,-35){\line(1,1){10}}
\put(0,-43){$c$}
\put(-10,25){$r-1$}
\end{picture}
\right] \nonumber \\
&=&- { b \over \sqrt{h} Z_c }
\left(
Z_{c\over 2}^2 \ln ( 4 Z_{c \over 2}^2 )
- Z_{h-c \over 2}^2 
\ln (4  Z_{ h- c \over 2}^2 )
\right)\, , \label{result1}
\end{eqnarray}
where $Z_a = \sin { a \pi \over h} $.
The divergent terms cancel each other and the total contribution
remains finite. 

With the help of various relations of the di-gamma
function and trigonometric function
one can prove that $\cal L$'s in (\ref{simply-power1}) 
and  (\ref{simply-power2}) coincide with the ones 
in (\ref{result1}).
Considering this as a non-trivial check, one can view this 
as a useful identity between di-gamma functions and trigonometric 
functions,
\begin{equation} \label{dseries}
b {\cal L}_i = \sum_{c= {\rm even}}^ {r-2}
<\Phi^c>_b\,   \bfe_i^c  \,.
\end{equation}
For example, we have for $i=1$, 
\begin{eqnarray} 
&&
\xi ( {1 \over h})  
+
\xi ( {2 \over h} ) - \xi ( {1 \over 2 } + {1 \over h}) 
-\xi ( {1 \over 2})
\nonumber \\
&&
= \sum_{c={\rm even}}^{r-2} 
8 \cos ({ c \pi \over h}) 
\left[
 \sin^2 (\frac {c\pi}{2h}) 
 \ln ( 4 \sin^2 ( \frac{c\pi}{2h}) ) 
-\cos^2 (\frac {c\pi}{2h}) 
 \ln ( 4 \cos^2 ( \frac{c\pi}{2h}) ) 
\right].
\end{eqnarray}

For the non-simply laced case,
the situation becomes more involved.
By expanding (\ref{moy}), 
one finds the following coefficients :
\begin{eqnarray}
{\cal K}&=&
\frac{{\bf d}}{h} \ln\left(\frac{2\mu}{l^2\mu'}\right)
  - \sum_{\val >0} \val^\vee\  
\ln \gamma \left( { \val \cdot \vrh^\vee \over h } 
\right)\,,\nonumber 
\\
{\cal L} &=& \frac{{\bf d}}{h} \left\{ \left( \frac{l^2}2 - 1 \right)
   [ 2 {\gamma_E } + \ln ( \pi \mu b^2 ) ]
   + (h - h^\vee) \ln\left(\frac{2\mu}{l^2\mu'}\right) \right\} \nonumber\\
&& - \int_0^\infty dt { 1 \over \sinh(h t)} 
  \left\{ 
   \coth(t) 
  \sum_{\val >0} \val\, 
\sinh (( h - 2 \val \cdot \vrh^\vee) t) 
\right.
\nonumber\\
&&\qquad\qquad\qquad\qquad
\left.
- { 2 \over h } 
 \sum_{\val >0} \val^\vee\,
\val \cdot ( h \vrh - h^\vee \vrh^\vee  )
 \cosh  (( h - 2 \val \cdot \vrh^\vee) t)
\right\}\,,\nonumber
\end{eqnarray}
where ${\gamma_E }=0.5772...$ is the Euler's number.

The explicit value for $B_r^{(1)}$ series takes the form :
\beqa
&& {\cal K}_1 
= { 2 \over h} 
\ln \left( {2 {\mu'} \over \mu}\right)
- \ln 2\ ;
\label{non-simply-L1}\\
&&{\cal K}_k = 
{2 \over h} 
\ln \left( {2 {\mu'} \over \mu}\right),
\quad 
k = 2, 3, \cdots, r-1\ ;  
\nonumber\\
&& {\cal K}_r =
- ( 1 - { 2\over h})\ln \left( {2 {\mu'} \over \mu}\right)
+ \ln 2\nonumber\ ;
\end{eqnarray}
and
\begin{eqnarray} \label{nonpert}
&& {\cal L}_1 
= {1 \over h} {\cal J} + {\cal I}_1  + {\Delta \cal I}_1\ ;
\label{non-simply-L2}\\
&& {\cal L}_r
= \left( {1 \over h }- {1 \over 2 }\right)
{\cal J} + {\cal I}_r  + {\Delta \cal I}_r\ ;
\nonumber\\
&&{\cal L} _k 
= {1 \over h }
{\cal J} + {\cal I}_k   + {\Delta \cal I}_k,
\ \ \   k = 2, 3, \cdots, r-1, \nonumber
\eeqa
where 
\begin{equation}
 {\cal J} = 
\left(
2 {\gamma_E } + \ln (\pi \mu b^2) 
+ {2 \over h} 
\ln \left( {{\mu'} \over 2 \mu}  \right) 
\right),
\end{equation} 
and 
\begin{eqnarray}
&&{\cal I}_1 = { 1\over 2h} 
\left\{
  \xi({ 1\over h})
+   \xi ( { 2 \over h}) 
- \xi ( { 1\over 2} + { 1 \over h}) 
- \xi ({ 1\over 2})
\right\}\ ;
\\
&&{\cal I}_r =  { 1\over 2h} 
\left\{
 2 \ \xi({ 1\over h})
- \xi ( { 2\over h} )
- \xi ({ 1\over 2})
\right\}\ ;
\nonumber\\
&&{\cal I}_k = { 1\over 2h} 
\left\{
\xi({ 2 k - 2 \over h})
+ 2 \ \xi ( {2k -1 \over h})
+   \xi ({2k \over h})
- \xi({k \over h})
\right.
\nonumber\\
&&\qquad \left.  \quad 
- \xi ({k-1 \over h})
- \xi( {1 \over 2}  +{k-1 \over h})
- \xi ({1 \over 2} + {k \over h})
\right\}\nonumber.
\end{eqnarray}
Note that ${\cal I}_k$'s ($k=1,\ldots,r-1$) are identical to 
${\cal L}_k$'s in (\ref{simply-power2}) for $D_{r+1}^{(1)}$ series.
${\Delta \cal I} $'s  are given by : 
\begin{eqnarray}
&&{\Delta \cal I}_1 =
{\Delta \cal I}_r   
= { 1\over 2h^2} 
\left\{
 - {2}\  \xi({ 1\over h})
+  {4}\  \xi ( { 2 \over h}) 
- {2} \ \xi ( { 1\over 2} + { 1 \over h}) 
\right\};
\\
&&{\Delta \cal I}_k = { 1\over 2h^2} 
\left\{
 (4-4k)\ \xi({ 2 k - 2 \over h})
+{4k} \ \xi ({2k \over h})
- {2k}\ \xi({k \over h})
\right.
\nonumber\\
&& \qquad\qquad \left. 
+(2k -2) \ \xi ({k-1 \over h})
+(2k -2) \ \xi( {1 \over 2}  +{k-1 \over h})
-{2k} \ \xi ({1 \over 2} + {k \over h})
\right\},\nonumber
\end{eqnarray}
and turn out to be identical to each other :
\begin{eqnarray}
&&{\Delta \cal I}_1 = {\Delta \cal I}_k = {\Delta \cal I}_r 
=   {4 \over h^2} \ln 2\ .
\end{eqnarray}

As noted for the simply laced case, ${\cal K}$ is identified with 
the classical value $\vph_{cl}$.
For $B_r^{(1)}$ series, 
\begin{equation}
b \bfe _i \cdot \vph_{cl} = 
\ln \big(  {\mu n_i \over \mu_{{\bfe}_i} } \big)
- {1 \over h} \sum_{j=0} ^r n_j 
\ln \big( {\mu n_j \over \mu_{{\bfe}_j}  }  \big)\,,
\end{equation}
which agrees with  $\cal K$ in (\ref{non-simply-L1}).

Beyond the classical result, however, renormalization 
should be carefully incorporated unlike in the simply laced case.
The classical mass eigenstate representation of $B_{r}^{(1)}$ 
is obtained by folding the one of $D_{r+1}^{(1)}$ (\ref{D-mass}),
\begin{eqnarray}
&&\bfe_1  = ( - l_1^1, -l_1^2, \cdots, -l_1^{r-1}, 1)
\label{eigen}\\
&&\bfe_k  = ( l_{k-1}^1- l_k^1, l_{k-1}^2-l_k^2, \cdots, 
l_{k-1}^{r-1} -l_k^{r-1},  0)
\ \ \mbox{ for } \ k =2, \cdots, r-1 
\nonumber\\
&&\bfe_r  = ( l_{r-1}^1, l_{r-1}^2, 
\cdots, l_{r-1}^{r-1}, 0)\nonumber
\end{eqnarray}
from which we obtain the one-loop contribution $<{\vph}>_b$ : 
\begin{eqnarray} \label{pert}
&&<\Phi^r >_b =0\ ;  
\nonumber\\
&&< \Phi^c >_b =  
{ b \over 4\, \sqrt{h}\,  Z_c^2}\, g_r\,  Z_{2c} 
\ \mbox { \rm when $c =$ odd } \le r-1 \,. \ ; 
\\
&&
<\Phi^c >_b = 
{ b \over 4 \,\sqrt{h}\,  Z_c^2} 
\left\{
{4 Z_c}
\left( g_{c \over 2}\, Z_{c\over 2}^2 
 - g_{ h-c \over 2}\, Z_{h-c \over 2}^2 
\right)
+ g_r \, Z_{2c}  
\right\}
\ \mbox { \rm when $c =$ even} \le r-1 \,, \nonumber
\end{eqnarray}
where $Z_a = \sin(a\pi/h)$ as is given in (\ref{result1}).  
$g_a$ is the Euclidean integration of the tad-pole diagram,
\begin{equation}
g_a \equiv \int { d^2 k \over (2\pi)^2 }
{ 1 \over k^2 + m_a^2 },
\end{equation}
where $m_a$ is the physical mass 
equivalent to $M_a$ in Appendix A up to this order of $b^2$.
Its explicit value is given by 
$m_a = 2 {\overline m}_0 \sin ( {\pi a \over h})$
for $a = 1, 2, \cdots , r-1$ and $m_r = {\overline m}_0$ with
${\overline m}_0^2 = 2^{2 + {2 \over h} } (\pi \mu b^2 )
 ( \mu' / \mu )^{2 \over h}$ .

Here, to evaluate the one-loop diagram 
we are using the normal ordering with respect to the free field
theory. In this scheme $g_a$ is given by  
\begin{eqnarray} \label{ga}
g_a &=& \frac{1}{4\pi} 
      \left( \ln ( {m_a  \over 2} ) ^2  + 2 {\gamma_E } \right) \nonumber\\
   &=&\left[ {\cal J} + \ln\left(\frac{m_a}{\overline{m}_0}\right)^2 
             + \frac2h \ln2\right].
\end{eqnarray}
Then, using the identity
\begin{equation}
\left( \sum_{c={\rm odd}}^{r-1} - \sum_{c={\rm even}}^{r-1} \right)
\csc^2(\frac{c\pi}h) \sin(\frac{2c\pi}h) \sin(\frac{2kc\pi}h) = \frac{k}{2h},
\qquad k=1,\ldots,r-1,
\end{equation}
we find that the ${\cal J}$ parts of (\ref{pert}) agree exactly with 
those of ({\ref{nonpert}), i.e. 
\[
b {\cal L}_i|_{{\cal J} \textrm{-part}} 
      = \sum_{c} \bfe_i^c  <\Phi^c>_b|_{{\cal J} \textrm{-part}}.   
\]
Furthermore, since the term $\ln(m_a/\overline{m}_0)^2$ in (\ref{ga}) 
reproduces ${\cal I}_k$'s which are the same as 
${\cal L}_k$'s in (\ref{simply-power2}) of $D_{r+1}^{(1)}$ series 
for $k=1,\ldots,r-1$, the agreement (\ref{dseries}) in 
$D_r^{(1)}$ case immediately implies
\[
b {\cal L}_i|_{{\cal I}\textrm{-part}} 
      = \sum_{c} \bfe_i^c  <\Phi^c>_b|_{{\cal I}\textrm{-part}}.
\]
Finally, $\Delta {\cal I}_k$ terms come from the last term $\frac2h\ln2$ in 
(\ref{ga}). This establishes the exact agreement between the perturbative 
and nonperturbative results for the $B_r^{(1)}$ case.

For the exceptional algebra $G_2^{(1)}$, we have 
\begin{eqnarray}
{\cal K}_1 &=& {1 \over 2} \ln({\mu' \over 3 \mu})
+ 2 \ln \gamma( {1 \over 6}) -4 \ln \gamma ({1 \over 3});
\\
{\cal K}_2 &=& - {1 \over 2} \ln({\mu' \over 3 \mu})
-  \ln \gamma( {1 \over 6}) + 2 \ln \gamma ({1 \over 3});
\nonumber\\
{\cal L }_1 & =& {1 \over 6} 
\left[ 4 \gamma_E  + \ln (( \pi \mu b^2 )
({ \pi \mu' b^2 \over 3})) \right] 
+ \frac12 \ln3 + \frac29 \ln2 ;
\nonumber\\
{\cal L }_2 &=&
-{1 \over 3} \left[  2 \gamma_E + {1 \over 2} \ln (( \pi \mu b^2 )
({ \pi \mu' b^2 \over 3})) \right] -\frac29\ln2 -\frac14 \ln3.
\nonumber 
\end{eqnarray}
On the other hand, the corresponding one-loop diagram is  given by : 
\begin{eqnarray}
<\Phi^1 >_b &=& \frac12\left[
\begin{picture}(50,50)(-25,0)
\put(0,0){\circle{40}}
\put(0,-20){\line(0,-1){10}}
\put(-5,-25){\line(1,-1){10}}
\put(-5,-35){\line(1,1){10}}
\put(0,-43){1}
\put(0,27){1}
\end{picture}
\right]
\end{eqnarray}
and 
\begin{eqnarray}
<\Phi^2 >_b &=& \frac12\left[
\begin{picture}(50,50)(-25,0)
\put(0,0){\circle{40}}
\put(0,-20){\line(0,-1){10}}
\put(-5,-25){\line(1,-1){10}}
\put(-5,-35){\line(1,1){10}}
\put(0,-43){2}
\put(0,27){2}
\end{picture}
+
\begin{picture}(50,50)(-25,0)
\put(0,0){\circle{40}}
\put(0,-20){\line(0,-1){10}}
\put(-5,-25){\line(1,-1){10}}
\put(-5,-35){\line(1,1){10}}
\put(0,-43){2}
\put(0,27){1}
\end{picture}
\right].
\end{eqnarray}
After explicit calculations as in the previous case, we find 
$<\Phi^2 >_b = b{\cal L}_2$ and 
$<\Phi^1 >_b  = {1 \over 2 } {\cal L}_1 + {\cal L }_2 $ which completes
the perturbative check for $G_2^{(1)}$.

\vspace{0.3cm}
\subsection{Perturbative checks of the composite operators $<\vph_a\vph_b>$}
\vspace{0.3cm}

>From the expression (\ref{Gafin}) and using (\ref{comp}) we have the 
VEV of composite operator, 
\begin{eqnarray}
G^{ab} &\equiv& << \vph^a \vph^b >> 
 \nonumber \\
&=&
- \delta^{ab} \sum_{i=1}^r
{n_i \over H(1 +b^2) }
 \ln (- \pi \mu_{{\bfe}_i} \gamma ( 1 + b^2 \bfe_i^2/2))
+ \int_0^\infty {dt \over t} 
[ \delta^{ab} e^{-2t} - {\cal F}^{ab}]\nonumber
\end{eqnarray}
where 
\begin{eqnarray}
{\cal F}^{ab} = b^2 t^2 
\sum_{\val >0}
\val^a \val^b 
{ \sinh (( 1 + b^2  \val ^2 /2)t)\,
\cosh \big(( (1 + b^2 )H - 2 b \val \cdot {\bf  Q })t\big)  
\over
\sinh(t)\ \ \sinh ( b^2 \val^2 t /2) 
\sinh ((1 + b^2 ) Ht) }\ .
\label{Fab}
\end{eqnarray}
These are in general rather complicated quantities to calculate perturbatively 
due to various divergences to be taken care of up to some finite part. 
For some combinations such as the relative value of the composite operator 
$G^{aa} - G^{rr}$ ($a = 1, \cdots, r-1 $), however, the propagators are 
renormalized with an over-all renormalization constant 
and therefore, most of the complications due to the renormalization scheme 
disappears. Therefore, such quantities provides an additional independent
check of the non-perturbative result in a simple way.
Since the perturbative calculation is done in the classical mass eigenstate 
representation, in this section we will use the mass eigenstate 
representation for $\val$ in (\ref{Fab}).

For $C_2^{(1)}$, the composite operator value is given by :   
\begin{eqnarray}
&&G^{12} = 0\ ;
\nonumber\\
&&G^{11} - G^{22}
= \int_0^\infty dt
{ 
b^2 t \, \sinh (1 +b^2 )t \, ( 4 \cosh (4 + 8b^2)t -4) 
\over 
\sinh t \, \sinh b^2 t \, \sinh (4 + 6b^2) t 
}\nonumber
\\
&&\qquad 
= \ln 2 + b^2 (0.79221\ldots) + {\cal O} (b^4)\,. \label{comp1}
\end{eqnarray}
The corresponding value is confirmed perturbatively: 
$ << \Phi^1 \Phi^2 >> = 0 $ 
since there is no vertex at all for this case.
The other one is given by :
\begin{eqnarray}
 \lefteqn{<< \Phi^1 \Phi^1 - \Phi^2 \Phi^2 >> } \nonumber \\
&&= 
\left[
\begin{picture}(50,40)(-25,0)
\put(0,0){\circle{40}}
\put(-5,-15){\line(1,-1){10}}
\put(-5,-25){\line(1,1){10}}
\put(0,25){1}
\end{picture}
-
\begin{picture}(50,40)(-25,0)
\put(0,0){\circle{40}}
\put(-5,-15){\line(1,-1){10}}
\put(-5,-25){\line(1,1){10}}
\put(0,25){2}
\end{picture}
\right]
+
\left[
\begin{picture}(50,40)(-25,0)
\put(0,0){\circle{40}}
\put(-20,0){\line(1,0){40}}
\put(-5,-15){\line(1,-1){10}}
\put(-5,-25){\line(1,1){10}}
\put(0,25){1}
\put(-15,-28){1}
\put(10,-28){1}
\put(0,3){2}
\end{picture}
- \frac12 \times
\begin{picture}(50,40)(-25,0)
\put(0,0){\circle{40}}
\put(-20,0){\line(1,0){40}}
\put(-5,-15){\line(1,-1){10}}
\put(-5,-25){\line(1,1){10}}
\put(0,25){1}
\put(-15,-28){2}
\put(10,-28){2}
\put(0,3){1}
\end{picture}
\right] +{\cal O} (b^4)
\nonumber\\
&&= 
\ln 2 + b^2 (0.79221\ldots) + {\cal O} (b^4)
\label{c2oneloop}
\end{eqnarray}
whose Feynman integration is done in the Appendix C.
This agrees with the non-perturbative results (\ref{comp1}).

For the case $B_3^{(1)}$, the non-perturbative result  gives : 
\begin{eqnarray}
&& G^{12} = G^{23} = 0\ ;
\nonumber\\
&&G^{11} - G^{33}
= -2 \int_0^\infty
{dt \over t} ({\cal F}^{11} - {\cal F}^{33}) 
=b^2 (-0.195326\ldots) + {\cal O} (b^4)\ ;\label{comp2}
\\
&&G^{22} - G^{33} 
= -2 \int_0^\infty
{dt \over t} ({\cal F}^{22} - {\cal F}^{33}) 
= -\ln 3 + b^2 (-0.321552\ldots) + {\cal O}(b^4)
\nonumber
\end{eqnarray}
where 
\begin{eqnarray}
&&{\cal F}^{11} - {\cal F}^{33} 
=
{
b^2 t^2 
\over 
\sinh (t) \, \sinh ((6 + 5 b^2 )t) 
}\ \times
\nonumber\\
&&\qquad
\left\{
{ \sinh ((1 +b^2)t) \over \sinh (b^2 t) }
\big(- \cosh ((4 +3 b^2)t)
+ 2 \cosh (b^2 t) - \cosh ((2+b^2)t)\big) 
\right. \ 
\nonumber\\
&&
\qquad
\left. \   
+ { \sinh ((1 +b^2/2 )t) \over 2 \sinh (b^2 t/2)  }
\big( \cosh ((4 +4 b^2)t) + \cosh ((2+b^2)t) - 2 \big)
\right\}\nonumber
 ;\\
&&{\cal F}^{22} - {\cal F}^{33} 
=
{
b^2 t^2 
\over 
\sinh (t) \, \sinh ((6 + 5 b^2 )t) 
}\ \times
\nonumber\\
&& \qquad
\left\{
{ \sinh ((1 +b^2)t) \over \sinh (b^2 t) }
\big( \cosh ((4 +3 b^2)t) - \cosh ((2+b^2)t) \big)
\right.\nonumber\\
&&\qquad \ 
+\left.
{ \sinh ((1 +b^2/2 )t) \over 2 \sinh( b^2 t/2)  }
\big( \cosh ((4 +4 b^2)t) + \cosh ((2+b^2)t) - 2 \big)
\right\}\nonumber \ .
\end{eqnarray}

The perturbative calculation gives the result, 
$ << \Phi^1 \Phi^3 >> = << \Phi^2 \Phi^3 >> = 0 $ 
since there is no vertex at all in this case.
The relative values of the composite operators are :
\begin{eqnarray}
<< \Phi^1 \Phi^1 - \Phi^3 \Phi^3 >> 
&=& \left[
\begin{picture}(50,40)(-25,0)
\put(0,0){\circle{40}}
\put(-5,-15){\line(1,-1){10}}
\put(-5,-25){\line(1,1){10}}
\put(0,25){1}
\end{picture}
+\begin{picture}(50,40)(-25,0)
\put(0,0){\circle{40}}
\put(-20,0){\line(1,0){40}}
\put(-5,-15){\line(1,-1){10}}
\put(-5,-25){\line(1,1){10}}
\put(0,25){1}
\put(-15,-28){1}
\put(10,-28){1}
\put(0,3){2}
\end{picture}
+
\frac12 \times
\begin{picture}(50,40)(-25,0)
\put(0,0){\circle{40}}
\put(-20,0){\line(1,0){40}}
\put(-5,-15){\line(1,-1){10}}
\put(-5,-25){\line(1,1){10}}
\put(0,25){3}
\put(-15,-28){1}
\put(10,-28){1}
\put(0,3){3}
\end{picture} \right] \nonumber \\
&& \qquad - \left[
\begin{picture}(50,40)(-25,0)
\put(0,0){\circle{40}}
\put(-5,-15){\line(1,-1){10}}
\put(-5,-25){\line(1,1){10}}
\put(0,25){3}
\end{picture}
+
\begin{picture}(50,40)(-25,0)
\put(0,0){\circle{40}}
\put(-20,0){\line(1,0){40}}
\put(-5,-15){\line(1,-1){10}}
\put(-5,-25){\line(1,1){10}}
\put(0,25){3}
\put(-15,-28){3}
\put(10,-28){3}
\put(0,3){1}
\end{picture}
+
\begin{picture}(50,40)(-25,0)
\put(0,0){\circle{40}}
\put(-20,0){\line(1,0){40}}
\put(-5,-15){\line(1,-1){10}}
\put(-5,-25){\line(1,1){10}}
\put(0,25){3}
\put(-15,-28){3}
\put(10,-28){3}
\put(0,3){2}
\end{picture} \right]
+ {\cal O} (b^4) \nonumber\\
&=&
b^2 (-0.195326\ldots) + {\cal O} (b^4) \ ;
\label{B3oneloop1}
\end{eqnarray}
\begin{eqnarray}
<< \Phi^2 \Phi^2 - \Phi^3 \Phi^3 >>
&=&
\begin{picture}(50,40)(-25,0)
\put(0,0){\circle{40}}
\put(-5,-15){\line(1,-1){10}}
\put(-5,-25){\line(1,1){10}}
\put(0,25){2}
\end{picture}
+ \frac12 \times \left\{
\begin{picture}(50,40)(-25,0)
\put(0,0){\circle{40}}
\put(-20,0){\line(1,0){40}}
\put(-5,-15){\line(1,-1){10}}
\put(-5,-25){\line(1,1){10}}
\put(0,25){1}
\put(-15,-28){2}
\put(10,-28){2}
\put(0,3){1}
\end{picture}
+
\begin{picture}(50,40)(-25,0)
\put(0,0){\circle{40}}
\put(-20,0){\line(1,0){40}}
\put(-5,-15){\line(1,-1){10}}
\put(-5,-25){\line(1,1){10}}
\put(0,25){3}
\put(-15,-28){2}
\put(10,-28){2}
\put(0,3){3}
\end{picture}
+
\begin{picture}(50,40)(-25,0)
\put(0,0){\circle{40}}
\put(-20,0){\line(1,0){40}}
\put(-5,-15){\line(1,-1){10}}
\put(-5,-25){\line(1,1){10}}
\put(0,25){2}
\put(-15,-28){2}
\put(10,-28){2}
\put(0,3){2}
\end{picture} \right\} \nonumber \\
&&\quad - \left[
\begin{picture}(50,40)(-25,0)
\put(0,0){\circle{40}}
\put(-5,-15){\line(1,-1){10}}
\put(-5,-25){\line(1,1){10}}
\put(0,25){3}
\end{picture}
 +
\begin{picture}(50,40)(-25,0)
\put(0,0){\circle{40}}
\put(-20,0){\line(1,0){40}}
\put(-5,-15){\line(1,-1){10}}
\put(-5,-25){\line(1,1){10}}
\put(0,25){3}
\put(-15,-28){3}
\put(10,-28){3}
\put(0,3){1}
\end{picture}
+
\begin{picture}(50,40)(-25,0)
\put(0,0){\circle{40}}
\put(-20,0){\line(1,0){40}}
\put(-5,-15){\line(1,-1){10}}
\put(-5,-25){\line(1,1){10}}
\put(0,25){3}
\put(-15,-28){3}
\put(10,-28){3}
\put(0,3){2}
\end{picture} \right]
+ {\cal O} (b^4) \nonumber\\
&=&
- \ln 3 + b^2 (-0.321552\ldots) + {\cal O} (b^4)
\label{B3oneloop2}
\end{eqnarray}
which exactly reproduces (\ref{comp2}). 

Similar check can be done for $G_2^{(1)}$. 
$G^{12}$ is not vanishing but is given by :
\begin{eqnarray}
G^{12} 
&&= -  {\sqrt{3}}\,  b^2 \sum_{\val >0} 
\big(({1 \over 2} \val_1 + \val_2 ) \cdot \val\big)\  
(\val_1 \cdot \val)
\nonumber\\ 
&&\times
\int { dt } 
\frac
{ t\ \sinh((1 + b^2 \alpha^2/2)t) \cosh((1 + b^2 ) H -2 b \val \cdot \vQ)t}
{\sinh(t) \sinh(b^2 \val^2 t /2) \sinh ((1 + b^2) H t) } 
\nonumber\\
&&=
{2 \over \sqrt{3}}\,  b^2 (0.0488314\ldots)  + O(b^4)  
\end{eqnarray}
whose value is also obtained from the perturbative diagram 
\begin{eqnarray}
<< \Phi^1 \Phi^2>> =
\frac12\times  
\begin{picture}(50,40)(-25,0)
\put(0,0){\circle{40}}
\put(-20,0){\line(1,0){40}}
\put(-5,-15){\line(1,-1){10}}
\put(-5,-25){\line(1,1){10}}
\put(0,25){1}
\put(-15,-28){1}
\put(10,-28){2}
\put(0,3){1}
\end{picture}\ + {\cal O} (b^4) \,. \label{Goneloop1}\\ \nonumber
\end{eqnarray}
\\
Finally, the relative value of the composite operators 
\begin{eqnarray}
&&G^{11} -G^{22}
=  - \int_{0}^{\infty} {dt \over \sinh (t) \sinh ((6 + 4 b^2)t)} 
\nonumber\\
&& \qquad\qquad
\quad
\left\{ \right.
\sinh((1 + b^2)t)
\left(
\cosh (2t) - \cosh(4t + 2b^2t) 
\right) 
\left.\right\}
\nonumber \\
&&
\qquad \qquad 
\quad  
+\
\sinh(t + {b^2 t \over 3}) 
\left( 2 \cosh (\frac{2b^2 t}{3}) - \cosh(4t + \frac{10b^2t}{3}) 
- \cosh(2t + \frac{4b^2t}{3}) \right)
\}
\nonumber\\
&&\qquad \qquad 
= {1 \over 2} \ln 3 + {2 \over 3} b^2(0.183165\ldots) + {\cal O}(b^4)
\end{eqnarray}
is reproduced by the  following perturbative diagrams,
\begin{eqnarray}
&&<< \Phi^1 \Phi^1 - \Phi^2 \Phi^2 >>
= \left[
\begin{picture}(50,40)(-25,0)
\put(0,0){\circle{40}}
\put(-5,-15){\line(1,-1){10}}
\put(-5,-25){\line(1,1){10}}
\put(0,25){1}
\end{picture}
+
\begin{picture}(50,40)(-25,0)
\put(0,0){\circle{40}}
\put(-20,0){\line(1,0){40}}
\put(-5,-15){\line(1,-1){10}}
\put(-5,-25){\line(1,1){10}}
\put(0,25){1}
\put(-15,-28){1}
\put(10,-28){1}
\put(0,3){2}
\end{picture}
+
\frac12 \times
\begin{picture}(50,40)(-25,0)
\put(0,0){\circle{40}}
\put(-20,0){\line(1,0){40}}
\put(-5,-15){\line(1,-1){10}}
\put(-5,-25){\line(1,1){10}}
\put(0,25){1}
\put(-15,-28){1}
\put(10,-28){1}
\put(0,3){1}
\end{picture} \right] \nonumber \\
&& \qquad \qquad   - \left[
\begin{picture}(50,40)(-25,0)
\put(0,0){\circle{40}}
\put(-5,-15){\line(1,-1){10}}
\put(-5,-25){\line(1,1){10}}
\put(0,25){2}
\end{picture}
+
\frac12 \times
\begin{picture}(50,40)(-25,0)
\put(0,0){\circle{40}}
\put(-20,0){\line(1,0){40}}
\put(-5,-15){\line(1,-1){10}}
\put(-5,-25){\line(1,1){10}}
\put(0,25){2}
\put(-15,-28){2}
\put(10,-28){2}
\put(0,3){2}
\end{picture}
+
\frac12 \times
\begin{picture}(50,40)(-25,0)
\put(0,0){\circle{40}}
\put(-20,0){\line(1,0){40}}
\put(-5,-15){\line(1,-1){10}}
\put(-5,-25){\line(1,1){10}}
\put(0,25){1}
\put(-15,-28){2}
\put(10,-28){2}
\put(0,3){1}
\end{picture} \right]
+ {\cal O} (b^4) \,. 
\label{Goneloop2}
\end{eqnarray}

It is straightforward to generalize the above perturbative calculation
to other remaining cases and to confirm the proposed  VEV.

\paragraph*{Acknowledgements}
P.B. is very grateful to V.A. Fateev for valuable discussions and
 for the hospitality of KIAS where part of this work was done. 
This work is supported in part by MOST 98-N6-01-01-A-05 (CA), 
and KOSEF 1999-2-112-001-5(CA,CR).
PB's work is supported in part by the EU under contract 
ERBFMRX CT960012 and Marie Curie fellowship HPMF-CT-1999-00094.\\

\section*{Appendix A}

\setcounter{equation}{0}

Differently to the simply laced case for which the mass ratios
correspond to the classical values, mass ratios for non-simply laced
case get quantum corrections \cite{DGZ,CDS}. The mass spectrum for the dual
cases remains the same with the change $b\rightarrow 1/b$, where the mass spectrum depends
only on one parameter $\mbar$ :
\beqa
B_r^{(1)}: &&  M_r=\mbar, \qquad  M_a =2\mbar\sin(\pi a/H),
                                \qquad a=1,2,\ldots,r-1 \nonumber\\
C_r^{(1)}: &&  M_a =2\mbar\sin(\pi a/H), \qquad a = 1,2,\ldots,r \nonumber\\
G_2^{(1)}: &&  M_1=\mbar, \qquad M_2 =2\mbar\cos(\pi(1/3 -1/H))\nonumber\\
F_4^{(1)}: &&  M_1=\mbar,\qquad M_2 =2\mbar\cos(\pi(1/3 -1/H)),\nonumber\\
           &&  M_3 =2\mbar \cos(\pi(1/6 -1/H)),\qquad
               M_4 = 2M_2 \cos(\pi/H)\,.   
\eeqa  
For non-simply laced Lie algebras, the Coxeter and dual Coxeter numbers are :
\beqa
&&  h_{B_r^{(1)}} =h_{(C_r^{(1)})}=2r, \qquad  h_{(B_r^{(1)})^{\vee}} = h_{A_{2r-1}^{(2)}}
=2r-1, \qquad h_{(C_r^{(1)})^{\vee}} = h_{D_{r+1}^{(2)}}=2(r+1), \nonumber\\
&&  h_{F_4^{(1)}} =12, \qquad h_{(F_4^{(1)})^{\vee}}=9, \qquad
\qquad h_{G_2^{(1)}}=6, \qquad h_{(G_2^{(1)})^{\vee}}= h_{D_4^{(3)}} =4.\nonumber
\eeqa\\

\section*{Appendix B : Notations}

\setcounter{equation}{0}
\beqa
\Phi_{\bf s}(A_2) &=& {\sqrt 2}\epsilon_2,\ \
\sqrt{3/2}\epsilon_1-1/{\sqrt 2}\epsilon_2; \nonumber\\
\Phi_{\bf s}(B_r) &=& \epsilon_i-\epsilon_{i+1}\ \ \ 1\leq i \leq r-1,\ \ \ \epsilon_r;\nonumber\\
\Phi_{\bf s}(C_r) &=& \epsilon_i-\epsilon_{i+1}\ \ \ 1\leq i \leq r-1,
\ \ 2\epsilon_r;\nonumber\\
\Phi_{\bf s}(D_r) &=& \epsilon_i-\epsilon_{i+1}\ \ \ 1\leq i \leq r-1,
\ \ \ \epsilon_r+\epsilon_{r-1};\nonumber \\
\Phi_{\bf s}(F_4) &=& \epsilon_i-\epsilon_{i+1}\ \ \ i\in\{2,3\},
\ \epsilon_4,\ \ \frac{1}{2}(\epsilon_1-\epsilon_2-\epsilon_3-\epsilon_4);\nonumber\\
\Phi_{\bf s}(G_2) &=& {\sqrt {2/3}}\epsilon_2,\ 
1/{\sqrt 2}\epsilon_1-{\sqrt {3/2}}\epsilon_2;\nonumber
\eeqa
and 
\beqa
&&{\overline \Phi_{\bf s}(C_r)}={\Phi_{\bf
s}(C_r)}|_{\epsilon_i\leftrightarrow \epsilon_{r+1-i}}\ ;\ \ \ \ \ 
 {\overline \Phi_{\bf s}(D_r)}={\Phi_{\bf
s}(D_r)}|_{\epsilon_i\leftrightarrow \epsilon_{r+1-i}}\ ;\nonumber\\
&&{\overline \Phi_{\bf s}(A_2)}={\Phi_{\bf
s}(A_2)}|_{\epsilon_1\leftrightarrow\epsilon_2}\ ;\ \ \ \ \ \ \ \ \ 
{\overline \Phi_{\bf s}(B_4)}={\Phi_{\bf
s}(B_4)}|_{\epsilon_i\leftrightarrow -\epsilon_i,\
i\in\{2,3,4\}}\ .\nonumber
\eeqa\\

\section*{Appendix C : Feynman integrals}

\setcounter{equation}{0}

The Feynman integration for $C_2^{(1)}$ in (\ref{c2oneloop}) is 
presented as the following. 
The lowest order diagrams (order of $b^0$) are represented as the 
Feynman integration :
\beqa
 <<\Phi^1 \Phi^1 - \Phi^2\Phi^2>>_0 
 = - 4 \pi \int \frac{d^2 p_E }{(2\pi)^2}
\left(
\frac{1}{p_E^2 + M_1^2} - \frac{1}{p_E^2 + M_2^2} \right)  
= \ln \frac{M_2^2}{M_1^2} = \ln2\,, 
\eeqa
where $p_E$ is the Euclidean momentum.
$M_i$ is the physical mass  and  its value at the integration 
is considered up to this appropriate perturbative order in $b$.
Since the wavefunction and mass renormalization is already done, 
the next-to leading order diagrams (order of $b^2$) are represented as :
\beqa
<<\Phi^1 \Phi^1 - \Phi^2\Phi^2>>_b 
=   (4 \pi)^2  \int \frac{d^2 p_E }{(2\pi)^2}
\left(\frac{32 \ I_{12} }{(p_E^2 + M_1^2)^2} 
- \frac{16\ I_{11}  }{p_E^2 + M_2^2} \right)  
= 0.79221\ldots
\eeqa
where 
\beqa
I_{ij}&&= \int\frac {d^2 k_E}{(2\pi)^2}
\frac{1}{(k_E^2 + M_i)^2} 
\frac{1}{(k_E+ p_E)^2 + M_j)^2} 
\nonumber\\
&&=\int_0^1 \frac{dx}{4\pi} 
\frac{1}{- x(1-x)p_E^2 + (1-x)M_i^2 + x M_j^2}
\eeqa

The Feynman integrations (\ref{B3oneloop1}) and (\ref{B3oneloop2})
of the next-to leading order for $B_3^{(1)}$ are given by :
\beqa
&&<< \Phi^1\Phi^1 - \Phi^3 \Phi^3>>_b 
 = (4 \pi)^2  \int \frac{d^2 p_E }{(2\pi)^2}
\left(\frac{( I_{33}  + 2\ I_{12})  }{(p_E^2 + M_1^2)^2} 
- \frac{(2 \ I_{13} + 2\ I_{23} ) }{(p_E^2 + M_3^2)^2} \right)  
\nonumber\\
&&\qquad\qquad\qquad
= - 0.195326\ldots
\nonumber\\
&& << \Phi^2\Phi^2 - \Phi^3 \Phi^3>>_b 
 = (4 \pi)^2  \int \frac{d^2 p_E }{(2\pi)^2}
\left(\frac{( I_{33} + I_{11}+  9\ I_{22}) }{(p_E^2 + M_2^2)^2} 
- \frac{(2 \ I_{13} + 2\ I_{23} ) }{(p_E^2 + M_3^2)^2} \right)  
\nonumber\\
&&\qquad\qquad\qquad 
= - 0.321552\ldots
\eeqa

The Feynman integrations (\ref{Goneloop1}) and (\ref{Goneloop2})
of the next-to leading order for $G_2^{(1)}$ are evaluated respectively as :
\beqa
&&<< \Phi^1\Phi^2>>_b 
 =  (4 \pi)^2 \frac{\ 2}{\sqrt3} 
\int \frac{d^2 p_E }{(2\pi)^2}
\left(
\frac{ I_{11} }{ (p_E^2 + M_1^2) (p_E^2 + M_2^2) } 
\right)
\nonumber\\
&&\qquad \qquad \qquad
=  \frac{\ 2}{\sqrt3} \times 0.0488314\ldots
\nonumber \\
&& << \Phi^1\Phi^1 - \Phi^2 \Phi^2>>_b 
 = (4 \pi)^2  \int \frac{d^2 p_E }{(2\pi)^2}
\left(\frac{( 2\ I_{12} + \frac 43 I_{11}) }{(p_E^2 + M_1^2)^2} 
- \frac{( 9 \ I_{12} + \ I_{11}) }{p_E^2 + M_2^2} \right)  
\nonumber\\
&&\qquad \qquad \qquad
= \frac 23 \times 0.183165\ldots
\eeqa


\begin{thebibliography}{10}
%
\bibitem{1}
A. Patashinskii and V. Pokrovskii, `` Fluctuation theory of phase
 transitions'', Oxford, Pergamon Press 1979.

\bibitem{2} M. Shifman, A. Vainstein and V. Zakharov, Nucl. Phys. {\bf B 147} (1979) 385.

\bibitem{3} Al. Zamolodchikov, Nucl. Phys.  {\bf B 348} (1991) 619.

\bibitem{4} S. Lukyanov and A. Zamolodchikov, Nucl. Phys.  {\bf B 493} (1997) 571. 
%
\bibitem{Fateev20}
V. A. Fateev, S. Lukyanov, A. Zamolodchikov and Al. B. Zamolodchikov, Phys. Lett. {\bf B 406} (1997) 83.
%
\bibitem{Zam0}
A. B. Zamolodchikov, Al. B. Zamolodchikov, Nucl.Phys. {\bf B 477} (1996) 577.
%
\bibitem{Fateev2}
V. A. Fateev, S. Lukyanov, A. Zamolodchikov and Al. B. Zamolodchikov, Nucl.Phys. {\bf B 516} (1998) 652.
%
\bibitem{8} A. LeClair, Phys. Lett.  {\bf B 230} (1989) 423;\\
            N. Rechetikhin and F. Smirnov, Commun. Math. Phys  {\bf 131} (1990) 157;\\
            D. Bernard and A. Leclair, Nucl. Phys.  {\bf B 340} (1990) 721;\\
            C. J. Efthimiou, Nucl. Phys. {\bf B 398} (1993) 697.

\bibitem{9} F. Smirnov, Int. J. Mod. Phys.  {\bf A 6} (1991) 1407.

\bibitem{10} R. Guida and N. Magnoli, Phys. Lett.  {\bf B 411} (1997)
127.
%
\bibitem{Bas}
P. Baseilhac, V.A. Fateev, Nucl.Phys. {\bf B 532} (1998) 567.
%
\bibitem{Fateev4}
V. A. Fateev, Mod. Phys. Lett. {\bf A 15} (2000) 259.
%
\bibitem{tr} C. Tracy and H. Widom, Commun. Math. Phys. {\bf 190} (1998)
697.
%
\bibitem{zN} H. J. de Vega and V.A. Fateev, J. Phys. {\bf A 25} (1992)
2693;\\
F. A. Smirnov, Int. J. Mod. Phys. {\bf A 6} (1991) 1407;\\
V. A. Fateev, Int. J. Mod. Phys. {\bf A 6} (1991) 2109.
%
\bibitem{vays}
I. Vaysburd, Phys. Lett. {\bf B 335} (1994) 161; Nucl. Phys. {\bf B 446}
(1995) 387.
%
\bibitem{muss} A. LeClair, A. W. W. Ludwig and G. Mussardo, Nucl.
Phys. {\bf B 512} (1998) 523.
%
\bibitem{zamfat}
V. A. Fateev and A. B. Zamolodchikov, Sov. Phys. JETP {\bf 63}
(1985) 215; Phys. Lett. {\bf A 92} (1992) 37.
%
\bibitem{coupled}
P. Baseilhac, Nucl. phys. {\bf B 594} (2001) 607.
%
\bibitem{non}
C. Ahn, P. Baseilhac, V.A. Fateev, C. Kim, C. Rim,  Phys. Lett. {\bf
B 481} (2000) 114.
%
\bibitem{hid}
C. Ahn, C. Kim, C. Rim, Nucl. Phys. {\bf B 556} (1999) 505.
%
\bibitem{kim}
C. Ahn, V.A. Fateev, C. Kim, C. Rim and B. Yang, Nucl. Phys. {\bf B 565}
(2000) 611.
%
\bibitem{YY} C. N. Yang and C. P. Yang, 
J. Math. Phys. {\bf 10} (1969) 1115.
%
\bibitem{Za} Al. B. Zamolodchikov, Nucl. Phys. {\bf B342} (1990) 695.
%
\bibitem{DGZ}
G. W. Delius, M. T. Grisaru and D. Zanon, Nucl. Phys. {\bf B 382}
(1992) 365.
%
\bibitem{CDS}
E. Corrigan, P. E. Dorey and R. Sasaki, Nucl. Phys. {\bf B 408}
(1993) 579.
%
\bibitem{pog1}
V.V.Mkhitaryan, R.H.Poghossian and T.A.Sedrakyan,
 ``Perturbation theory in radial quantization approach and the
 expectation values of exponential fields in sine-Gordon model'', 
hep-th/9910128.
%
\bibitem{pog2}
R.H.Poghossian, Nucl.Phys. {\bf B 570} (2000) 506.
%
\bibitem{Hol}
T. Hollowood and P. Mansfield, Phys. Lett. {\bf B 226} (1989) 73.
%
\bibitem{Curt}
T. Curtright and C. Thorn, Phys. Rev. Lett. {\bf 48} (1982) 1309;\\
E. Braaten, T. Curtright and C. Thorn, Phys. Lett. {\bf B 118} (1982)
115; Ann. Phys. {\bf 147} (1983) 365.

\bibitem{Neveu}
J. L. Gervais and A. Neveu, Nucl. Phys. {\bf B 238} (1984) 125; 
{\bf B 238} (1984) 396; {\bf B 257}[FS14] (1985) 59.
%
\bibitem{FL} V. Fateev and S. Lukyanov, Sov. Sci. Rev. {\bf A212} 
(Physics) (1990) 212.
%
\bibitem{BS} P. Bouwknegt and  K. Schoutens, "W-symmetry" Singapore, 
World Scientific (1995).
%
\bibitem{Zam1}
A. B. Zamolodchikov, Adv. Stud. in Pure Math. {\bf 19} (1989) 641.
%
\bibitem{bernard}
D. Bernard and A. Leclair, Commun. Math. Phys. {\bf 142} (1991) 99.
%
\bibitem{mumass} Al. B. Zamolodchikov, Int. J. Mod. Phys. {\bf A10}
(1995) 1125.
%
\bibitem{fateev} V. A. Fateev, Phys. Lett. {\bf B324} (1994) 45.
%
\bibitem{moi}
P. Baseilhac, JHEP Proceedings of the 4-th annual TMR conference
``Nonperturbative quantum effects 2000'', hep-th/0009245. 
%
\bibitem{destri} C. Destri and H.J. de Vega, Nucl. Phys. {\bf B 358} (1991) 251. 
\end{thebibliography}
\end{document}